 \documentclass[preprint2]{aastex}

\slugcomment{To appear in the Astrophysical Journal}

\shorttitle{Large-Scale Variability in Wolf-Rayet Winds II}
\shortauthors{Chen\'e \& St-Louis}

\begin{document}

\title{A Systematic Search for Corotating Interaction Regions in Apparently Single Galactic Wolf-Rayet Stars. II. A Global View of the Wind Variability.}

\author{A.-N. Chen\'e}
\affil{Departamento de Astronom\'ia, Casilla 160-C, Universidad de Concepci\'on, Chile \\
    Departamento de F\'isica y Astronom\'ia, Facultad de Ciencias, Universidad de Valpara\'iso, Av. Gran Breta\~na 1111, Playa Ancha, Casilla 5030, Valpara\'iso, Chile \\
Canadian Gemini Office, Herzberg Institute of Astrophysics, 5071, West Saanich Road, Victoria (BC), Canada V9E 2E7}
 \email{achene@astro-udec.cl}

\and

\author{N. St-Louis} \affil{D\'{e}partement de Physique and Centre de
Recherche en Astrophysique du Qu\'ebec, Universit\'{e} de
Montr\'{e}al, C.P. 6128, Succ. Centre-Ville, Montr\'{e}al, Qu\'{e}bec,
Canada H3C 3J7} \email{stlouis@astro.umontreal.ca}

\vspace{1cm}

\begin{abstract}
This study is the second part of a survey searching for large-scale spectroscopic variability in apparently single Wolf-Rayet (WR) stars. In a previous paper (Paper~I), we described and characterized the spectroscopic variability level of 25 WR stars observable from the northern hemisphere and found 3 new candidates presenting large-scale wind variability, potentially originating from large-scale structures named Co-rotating Interaction Regions (CIRs). In this second paper, we discuss an additional 39 stars observable from the southern hemisphere. For each star in our sample, we obtained 4--5 high-resolution spectra with a signal-to-noise ratio of $\sim$100 and determined its variability level using the approach described in Paper~I. In total, 10 new stars are found to show large-scale spectral variability of which 7 present CIR-type changes (WR\,8, WR\,44, WR55, WR\,58, WR\,61, WR\,63, WR\,100). Of the remaining stars, 20 were found to show small-amplitude changes and 9 were found to show no spectral variability as far as can be concluded from the data in hand.

Also, we discuss the spectroscopic variability level of all single galactic WR stars that are brighter than $v\sim12.5$, and some WR stars with $12.5< v \leq 13.5$; i.e. all the stars presented in our two papers and 4 more stars for which spectra have already been published in the literature. We find that 23/68 stars (33.8\%) present large-scale variability, but only 12/54 stars ($\sim$22.1\%) are potentially of CIR-type. Also, we find 31/68 stars (45.6\%) that only show small-scale variability, most likely due to clumping in the wind. Finally, no spectral variability is detected based on the data in hand for 14/68 (20.6\%) stars. Interestingly, the variability with the highest amplitude also have the widest mean velocity dispersion.
\end{abstract}


\section{Introduction}

Up to date, no reliable direct observations of Wolf-Rayet (WR) rotation rates have been obtained. In fact, since the spectrum of WR stars originates exclusively in their hot and dense wind, the surface cannot be seen and no ``classical'' photospheric line can be observed. This renders the direct determination of rotational velocities of WR stars using Doppler broadening of photospheric lines impossible. However, in some cases the rotation period can be deduced for single WR stars presenting periodic wind variability. When present in a WR wind, this type of variability could be due to the presence of Co-rotating Interaction Regions (CIRs) generated by perturbations at the base of the wind which propagate through it and are carried around by rotation. Although the detected period is not necessarily directly the rotation period of the star, there is likely a link between both periods that can be untangled with a proper monitoring and modeling of spectral lines \citep[e.g. the case of the B supergiant star discussed in][]{Lo08}.

Only two WR stars were known to show CIR-type variability; i.e. WR\,6 and WR\,134 which have a unique photometric period also present in the line-profile variability \citep{Fir,Mor1,Mor2}. A binary scenario has also been proposed to explain the periodic variations, but no clear evidence has ever been found for the presence of a companion.  In order to identify more candidates, we \citep[in ][hereafter referred to as Paper~I]{Stlouis:2009} have obtained 4--5 spectra well spaced in time of 25 northern, galactic WR stars which are thought to be single and brighter than $v\sim$ 12.5. Those observations allowed us to establish a way to group the WR stars in terms of the strength of their spectral variability, i.e. stars that do not show variability based on the data in hand (NV), stars that show small-scale variability (SSV) and stars that show large-scale variability (LSV). Our CIR-type candidates are taken among the LSV group. However, we did not consider that all LSVs are good CIR-type variable candidates. Indeed, we found that WN8 stars show variability that appears  different in nature from those we are attempting to identify (see Paper~I), and might have a different origin \citep{Ant,Mar}. WC9 stars were also excluded since many WC stars of this sub-type have been shown to be binaries \citep{Hac76,Wil78} and present spectral variations that, without proper monitoring, might look like CIR-type changes, but in reality be binary-related. These two classes of stars deserve a more thorough and dedicated study. Finally, follow-up observations of one candidate, WR\,1, have recently confirmed the presence of CIR-type variability in that star \citep{Che10}.

In this paper, we complete the survey initiated in Paper~I of all WR stars in the catalogue of \citet{Huc01} brighter than $v\sim$12.5, and we add stars down to a magnitude of $v\sim13.5$ for a total of 39 apparently single southern, galactic WR stars. We have not included any WN8 stars in this southern sample for the reason mentioned above. Also, some stars brighter than the chosen magnitude limit have been excluded from our list, i.e. WR\,25 for which an orbital period has since been identified \citep{Gamen:2006}, WR\,94 which turned out to be a visual binary and WR\,7 and WR\,60, for which we unfortunately did not get enough spectra to investigate their variability. Our targets are listed in Table~\ref{Targets} where the stars' name, spectral type, RA, DEC, $v$ magnitudes and terminal velocity, $v_\infty$ (when available in the literature), are provided. The last column indicates the stars' variability status (see below).

Section~\ref{Obs} summarizes the observations. The search for variability is described in Section~\ref{VarSer} and a discussion of the results for the new sample of stars and a summary of the spectroscopic variability level of all the Galactic WR stars in our global sample are presented in Section~\ref{Disc}.

\section{Observations}\label{Obs}

We carried out spectroscopic observations using the 1.5m-telescope of the Cerro Tololo Inter-American Observatory (CTIO) during 14 nights between 2004 June 28 and July 12. As for the stars discussed in Paper~I, we obtained 4--5 spectra per target with the long-slit spectrograph. We have reached a typical signal-to-noise ratio of 100 over a spectral range of 5200--6000 \AA\ and with a resolution of 2.3 \AA\ (3 pix.).

The data reduction was performed using the standard {\sc iraf} software packages. First, a bias frame was subtracted, then each image was divided by a flat field. After extracting the spectra, the wavelength calibration was carried out using a He-Ar calibration lamp. As our spectra are not photometrically calibrated, the final step of our data reduction procedure was to rectify the spectra. To do so, we have fitted a low-order Legendre polynomial to wavelength regions with no strong spectral lines and divided our spectra by the fitted curve.

\subsection{Choice of lines}

The spectral interval chosen for this work is different from that adopted for the stars presented in Paper~I. Actually, in the previous work, the initial wavelength interval ranged from 4500 to 6000 \AA. However, due to problems with the wavelength calibration, the reddest half of the spectrum had unfortunately to be excluded and the useful spectral range was $\lambda\lambda$=4500--5200\AA. For WN stars, we were therefore forced to limit our study to the He{\sc ii}$\lambda$4686 and He{\sc ii}$\lambda$4860 lines, which are very similar. For WC stars, except for the very strong C{\sc iii}$\lambda$4650 line, which is often blended with a few faint lines, only weak lines were available. Nevertheless, the above-mentioned transitions were able to provide us with a diagnostic of the line-profile variability level in most cases. However, for this study, we prefered to choose a different wavelength interval in order to monitor lines that are potentially of greater interest (i.e. formed at different radii, more sensitive to density variability, etc...).

In WN spectra, there are two relatively strong isolated lines which are interesting to study within the 800\AA-wide spectral interval available at the CTIO-1.5m telescope. These are He{\sc ii}$\lambda$5411, often among the strongest emission lines in the optical band, and He{\sc i}$\lambda$5876, a flat-topped line only slightly perturbed by two NaD interstellar lines. These two emission lines of different ionization states do not have the same shape, which indicates that they are formed in different velocity regimes of the wind, i.e. at different radii. \citet{Sc95} present a plot of the  ionization stratification of their model of the WN4 star WR\,6; in general He{\sc i} lines are found to form at higher velocity than He{\sc ii} lines. For WC spectra, we chose C{\sc iii}$\lambda$5696, a flat-topped line particularly sensitive to density variations, and C{\sc iv}$\lambda\lambda$5802/12, a doublet that is unfortunately blended in some cases with several other lines. For WN/WC stars, we looked at the He{\sc ii}$\lambda$5411 and C{\sc iv}$\lambda\lambda$5802/12 lines. The two lines are also formed in different (although partly overlapping) velocity regions of the wind. \citet{De00} for example, present line-formation regions for the WC8 star WR\,135; they find that the C{\sc iii}$\lambda$5696 line peaks at a distance about twice as far as the C{\sc iv}$\lambda$5802/12 doublet.

\section{Variability Search}\label{VarSer}

To search for significant line-profile variability of the targets, the Temporal Variance Spectrum (TVS) of each dataset was calculated using the formalism of \citet{Ful} and the quantity $\Sigma_j(99\%)=\sqrt{\frac{(TVS)_j}{\sigma_0^2 \chi^2_{N-1}(99\%)}}$ was obtained, where $\sigma_0$ is the reciprocal of the rms of the noise level in the continuum in a timeseries of N spectra. The value of $\Sigma_j$(99\%) quantifies the level of variability at each wavelength: a spectrum that reaches a value of $n$ varies with an amplitude $n$ times higher than the variability measured in the continuum (which is assumed to be pure noise) with a confidence level of 99\%. The spectrum of a given star is considered significantly variable at a given wavelength $j$ if the value of $\Sigma_j$(99\%) is significantly greater than 1. When a line is identified as significantly variable, it is possible to calculate the amplitude of its variability relative to its intensity. To do so, we have calculated in Paper~I for each wavelength $j$ a modified TVS$^{1/2}_j$ as defined in \citet{Che} and divided it by the line flux $(\bar{S_j}-1)$, where $\bar{S_j}$ is the weighted mean flux at wavelength $j$. This ratio is named the $\sigma$-spectrum. One should note that the calculation of $\sigma$ does not take into account instrumental variations due to the noise level. Thus, when the variation level of a given line is too close to the noise, which is the case for weaker lines, $\sigma$ is artificially high. That is why we manually set $\sigma_j$=0 when the variability at a given wavelength $j$ is not clearly significant according to the value of $\Sigma_j$(99\%) or when the line has a relative intensity lower than 1.3, which is the intensity limit of a spectral line from which we can investigate the variability for the present dataset. Using the $\Sigma$(99\%) spectrum and the $\sigma$-spectrum, three main categories were defined in Paper~I: stars showing no profile variability (NV~: $\Sigma$(99\%) spectrum $\lesssim$ 1), stars showing small-scale profile variability (SSV~: $\sigma$ $\leq$ 5 \%) and stars showing large-scale profile variability (LSV~: $\sigma$ $>$ 5\%).  One has to note that in the case of many WC stars, the global line intensity seems to vary from spectrum to spectrum. These changes are specially big in the case of the broad, strong emission lines (with relative intensity greater than $\sim\,5$), probably due to the difficulty to fit the continuum in these regions. In these cases, we have normalized the line intensity in order to isolate the variability cause by bumps.

\begin{figure}[ht]
  \centerline{\plotone{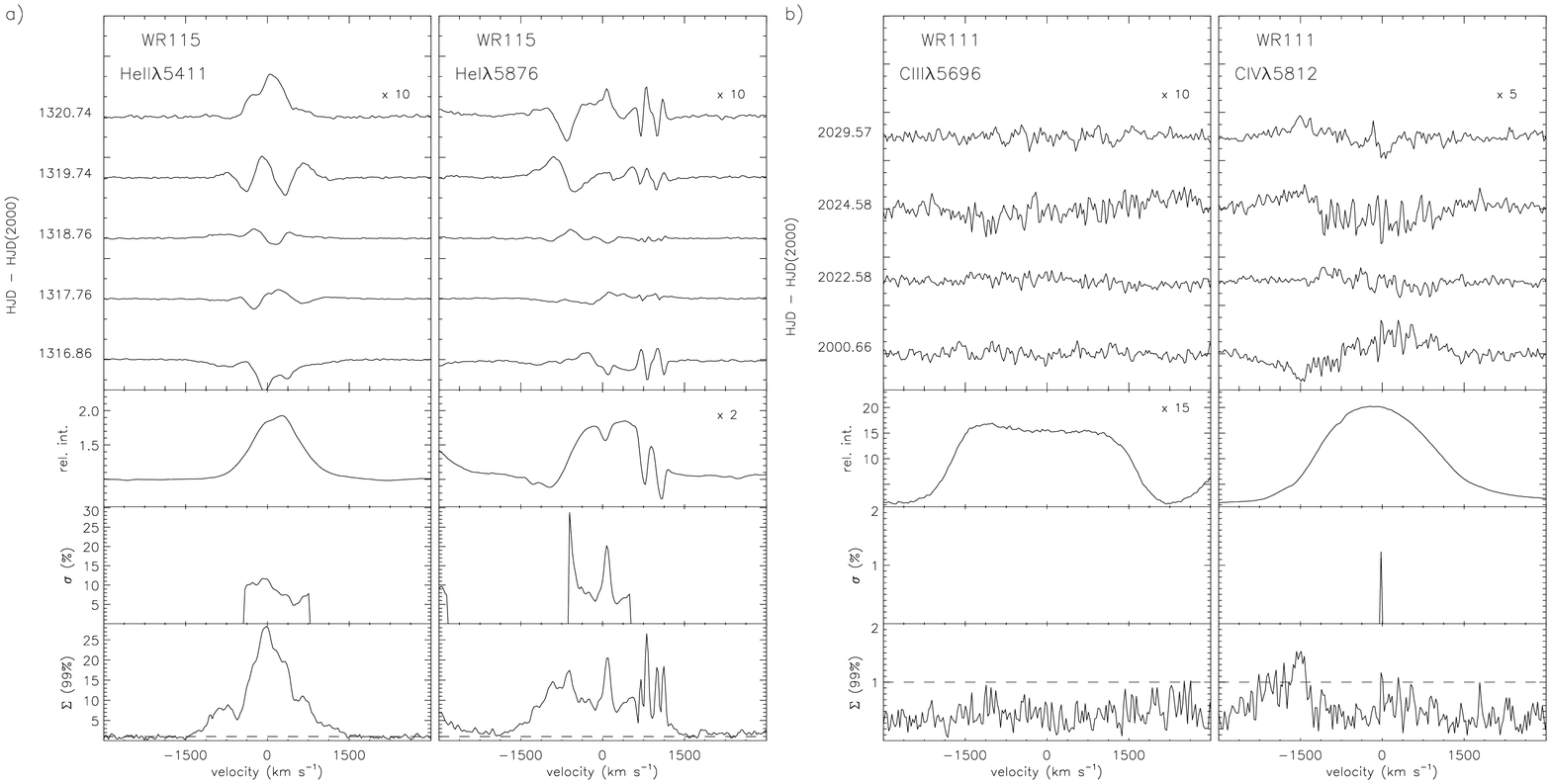}}
  \caption{a) Top: Montage of the He{\sc ii}$\lambda$5411 (left) and
  He{\sc i}$\lambda$5876 (right) residuals (individual spectra --
  mean) for WR\,115 (WN6). For both cases, the scale factor of the
  ordinate is indicated in the top right-hand corner of the plot. HJD
  - HJD(2000) is indicated on the {\it y}-axis. Second from top: mean
  spectrum. Second from bottom: $\sigma$-spectrum. Bottom:
  $\Sigma$~(99\%) spectrum. b) Same as a) for WR\,111 (WC5).}
 \label{NordSud}
\end{figure}

For this study, we do not use the same spectral lines as in Paper~I and one may wonder if this could affect our variability diagnostic. In fact, it has been shown in Paper~I that the $\sigma$-value does not change significantly from line to line, as long as the noise level is not of the same order or higher than the amplitude of the variability. To demonstrate that this is also true for the lines we have chosen in this study, we re-observed two targets already presented in Paper~I. In Figure~\ref{NordSud}(a), we present in the upper panel, a montage of residuals obtained by subtracting the mean from individual spectra for WR\,115 (WN6). These observations were obtained with the OSIS spectrograph at the Canada-France-Hawaii telescope (CFHT) during 6 nights in 2003 August ($\Delta\lambda$ = 1.5 \AA\ (3 pix.), $\lambda\lambda$ = 4400-6100 \AA, S/N =150). Using the $\sigma$-spectrum shown in the middle panel of Figure~\ref{NordSud}(a) and the same criteria established in Paper~I, we find that, as in Paper~I, WR\,115 is designated as LSV.  This is also the case for WC stars, as shown in Figure~\ref{NordSud}(b), where we present the montage of residuals of WR\,111 (WC5) obtained with the long-slit spectrograph of the Observatoire du mont M\'egantic (OMM) in 2005 and 2006 ($\Delta\lambda$ = 1.6 \AA\ (3 pix.), $\lambda\lambda$ = 4500-6100 \AA, S/N =100). Indeed, this figure shows that, as in Paper~I, WR\,111 is designated as NV (or on the verge of being classified as SSV when looking at the C{\sc iv}$\lambda\lambda$5802/12 line). Moreover, for the strongest lines, when variability is detected, the diagnostic deduced from the $\sigma$-spectrum does not depend on the S/N of the spectra.

Interestingly, Figure~\ref{NordSud} also shows that even when the lines are formed at different radii in the stellar wind and in different physical conditions, the observed variability does not seem to change significantly in amplitude. Of course, only a better monitoring would allow us to determine if the spectral features causing the variability have different shapes depending on the emission line and/or if time delays exist  between the variations observed in different lines.

\begin{figure}[tbp]
  \centerline{\plotone{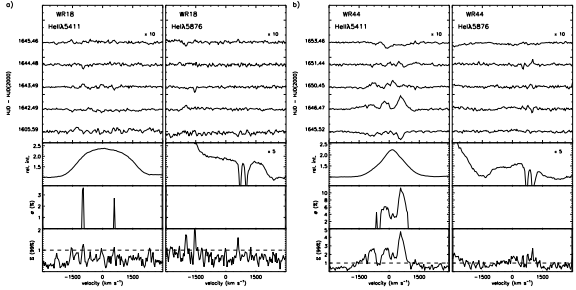}}
  \caption{Same as Figure~\ref{NordSud}(a) for a) WR\,18 (WN4) and b) WR\,44 (WN4+OB?).}
  \label{fig2S}
\end{figure}

\begin{figure}[tbp]
  \centerline{\plotone{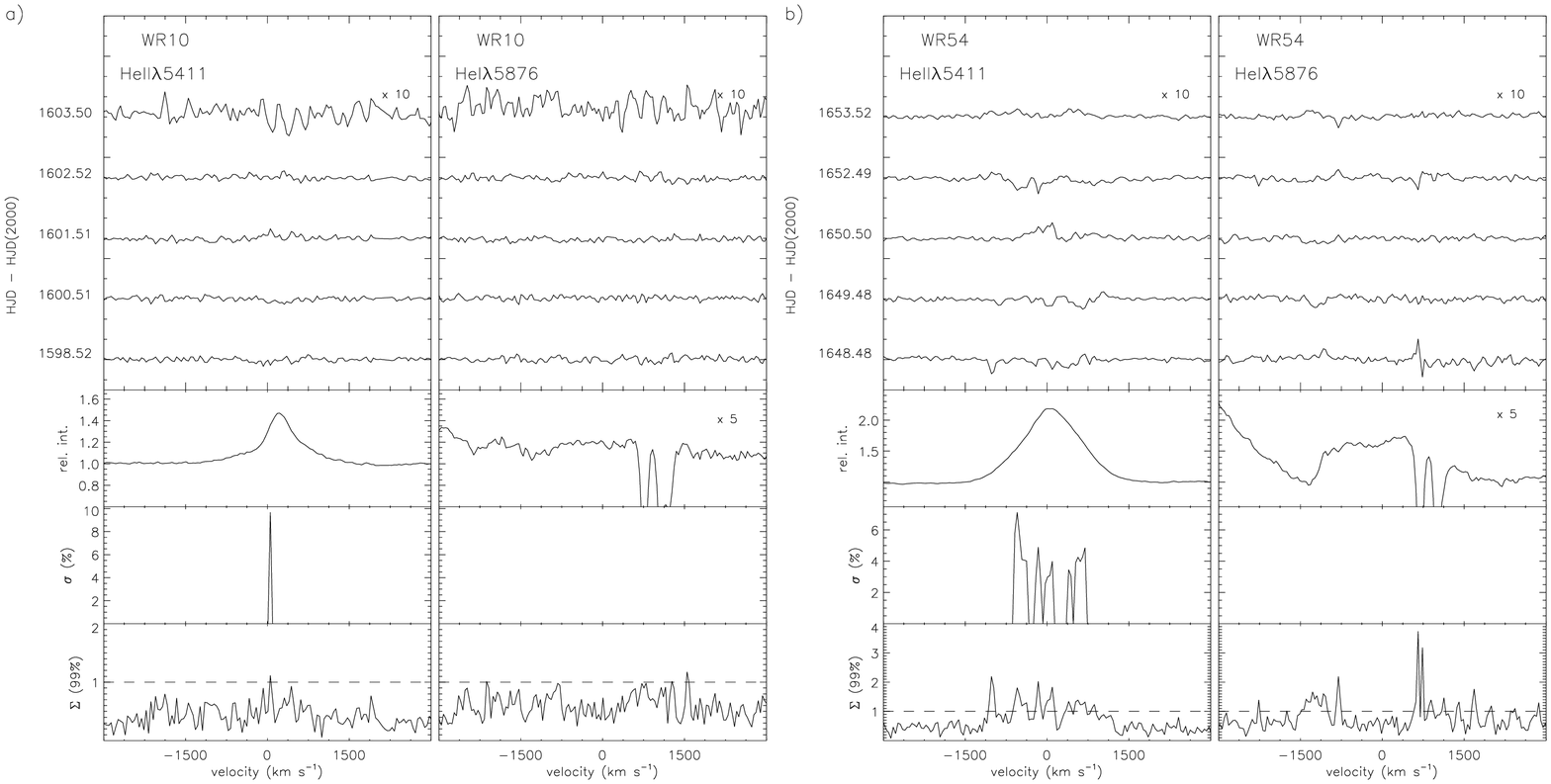}}
  \caption{Same as Figure~\ref{NordSud}(a) for a) WR\,10 (WN5ha) and b) WR\,54 (WN5).}
  \label{fig3S}
\end{figure}

\begin{figure}[tbp]
  \centerline{\plotone{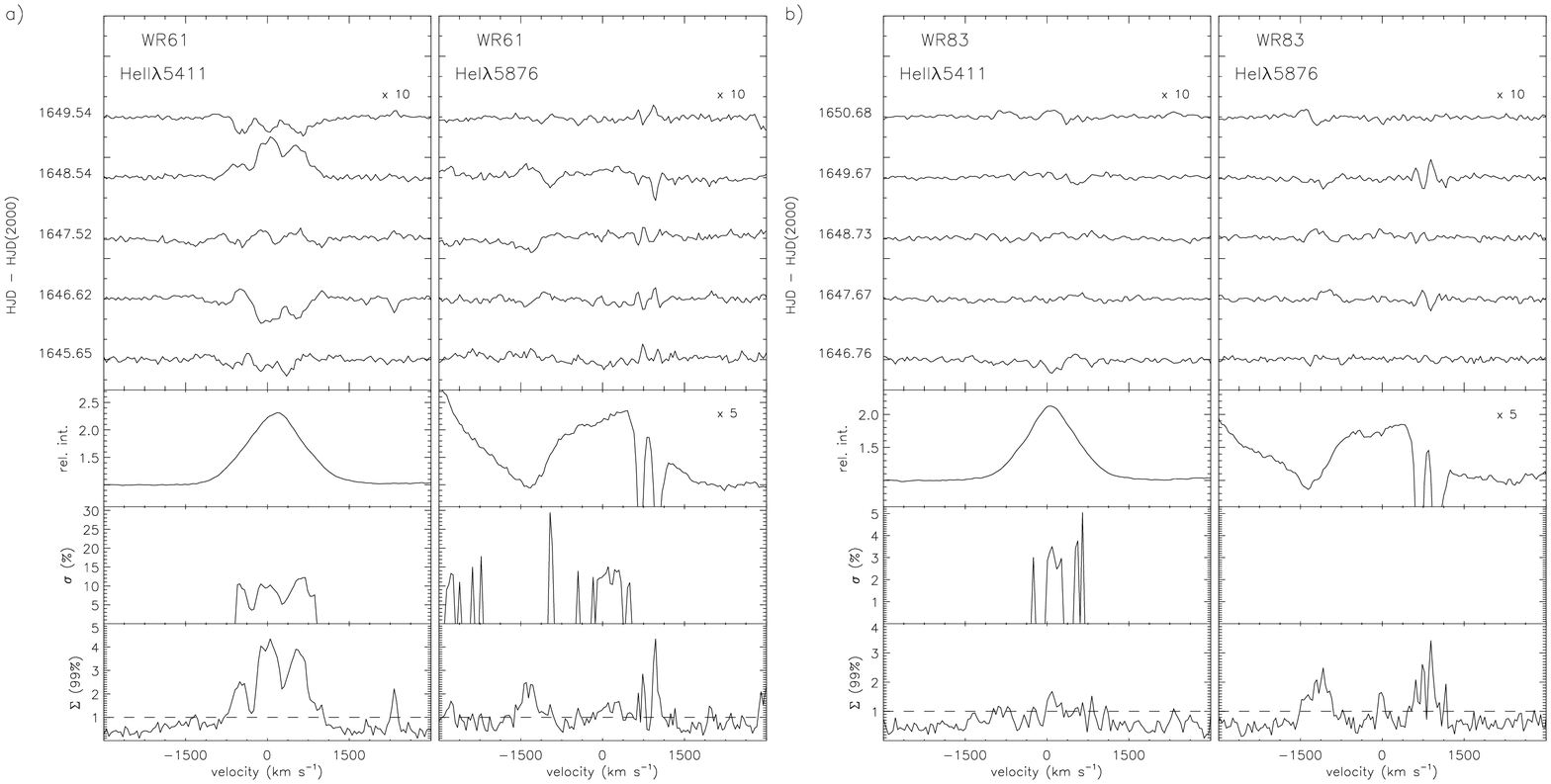}}
  \caption{Same as Figure~\ref{NordSud}(a) for a) WR\,61 (WN5) and b) WR\,83 (WN5).}
  \label{fig4S}
\end{figure}

\begin{figure}[tbp]
  \centerline{\plotone{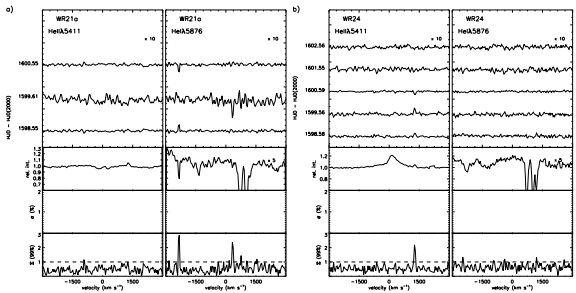}}
  \caption{Same as Figure~\ref{NordSud}(a) for a) WR\,21a (WN6+O/a) and b) WR\,24 (WN6ha).}
  \label{fig5S}
\end{figure}

\begin{figure}[tbp]
  \centerline{\plotone{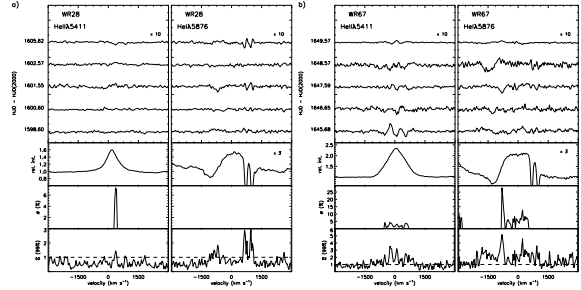}}
  \caption{Same as Figure~\ref{NordSud}(a) for a) WR\,28 (WN6(h)+OB?) and b) WR\,67 (WN6+OB?).}
  \label{fig6S}
\end{figure}

\begin{figure}[tbp]
  \centerline{\plotone{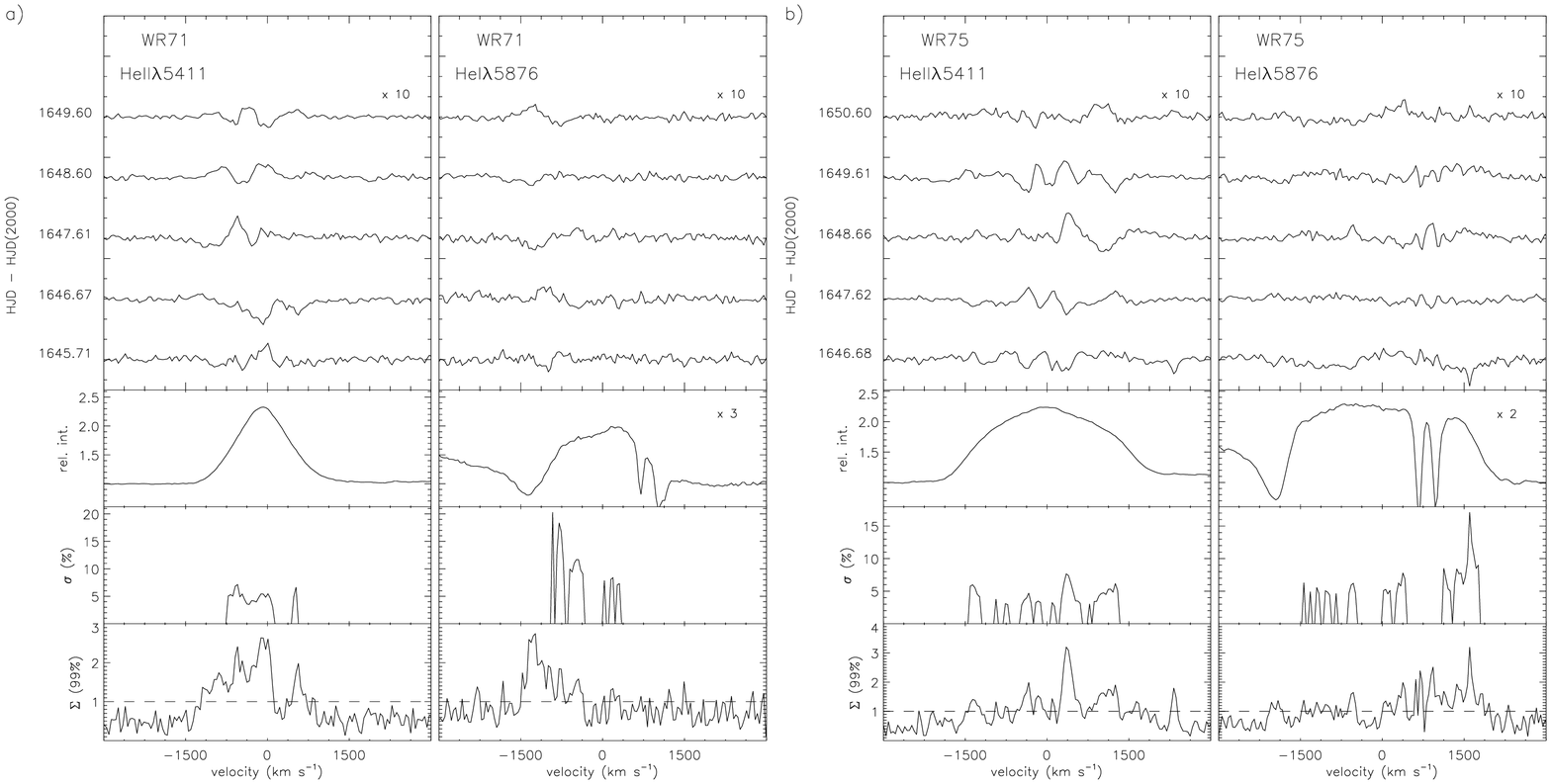}}
  \caption{Same as Figure~\ref{NordSud}(a) for a) WR\,71 (WN6+OB?) and b) WR\,75 (WN6).}
  \label{fig7S}
\end{figure}

\begin{figure}[tbp]
  \centerline{\plotone{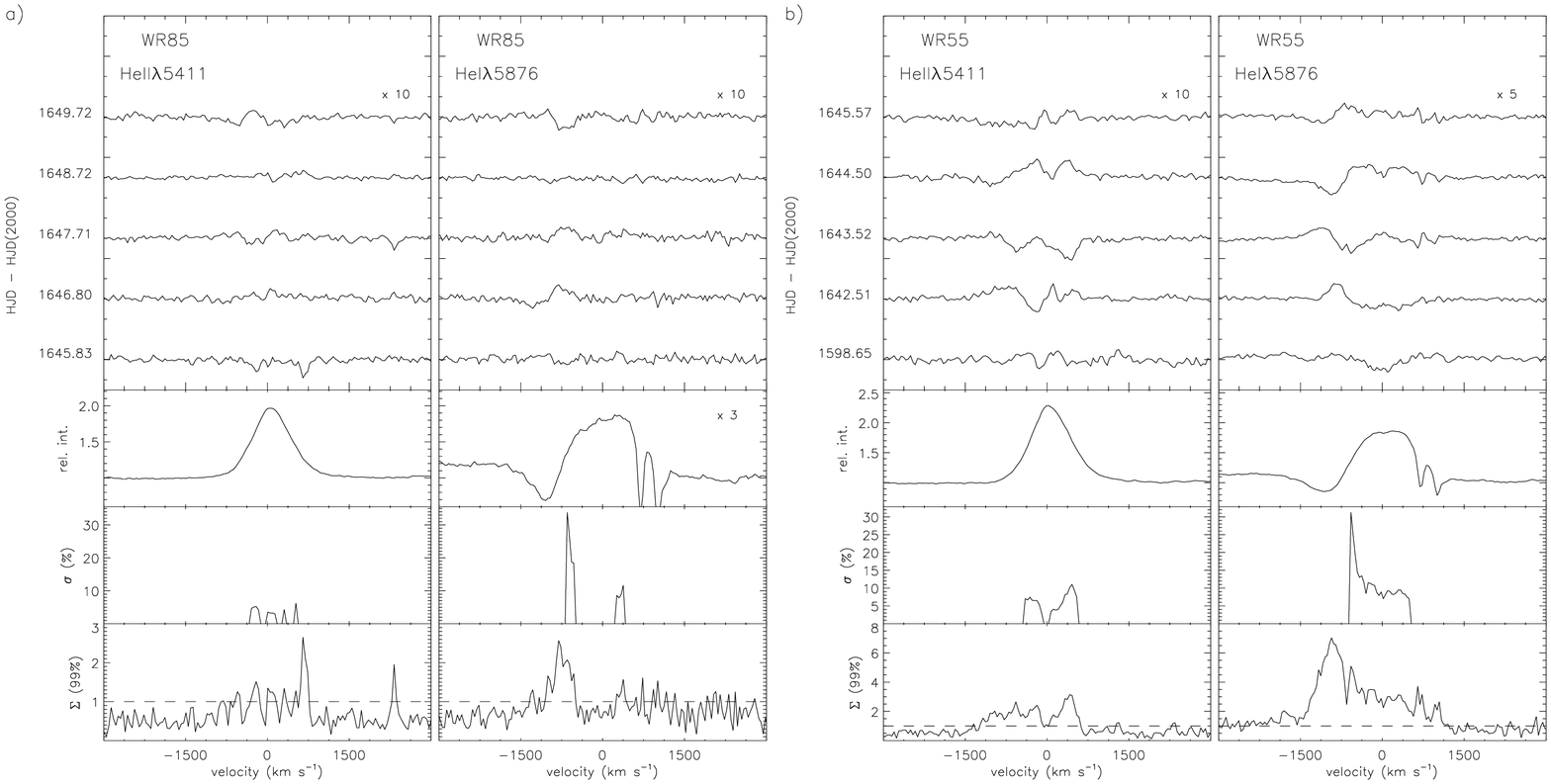}}
  \caption{Same as Figure~\ref{NordSud}(a) for a) WR\,85 (WN6h+OB?) and b) WR\,55 (WN7).}
  \label{fig8S}
\end{figure}

\begin{figure}[tbp]
  \centerline{\plotone{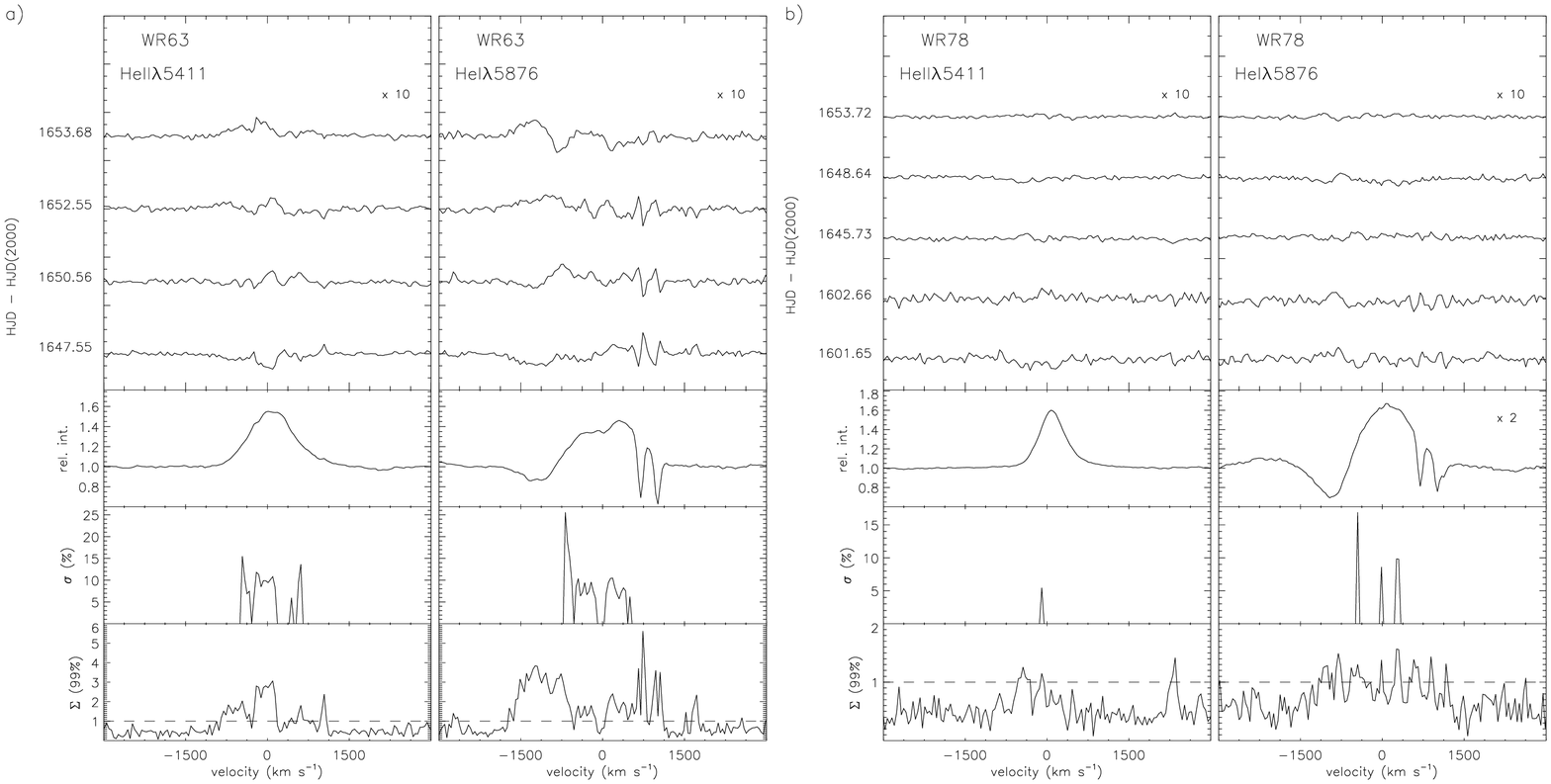}}
  \caption{Same as Figure~\ref{NordSud}(a) for a) WR\,63 (WN7+OB?) and b) WR\,78 (WN7h).}
  \label{fig9S}
\end{figure}

\begin{figure}[tbp]
  \centerline{\plotone{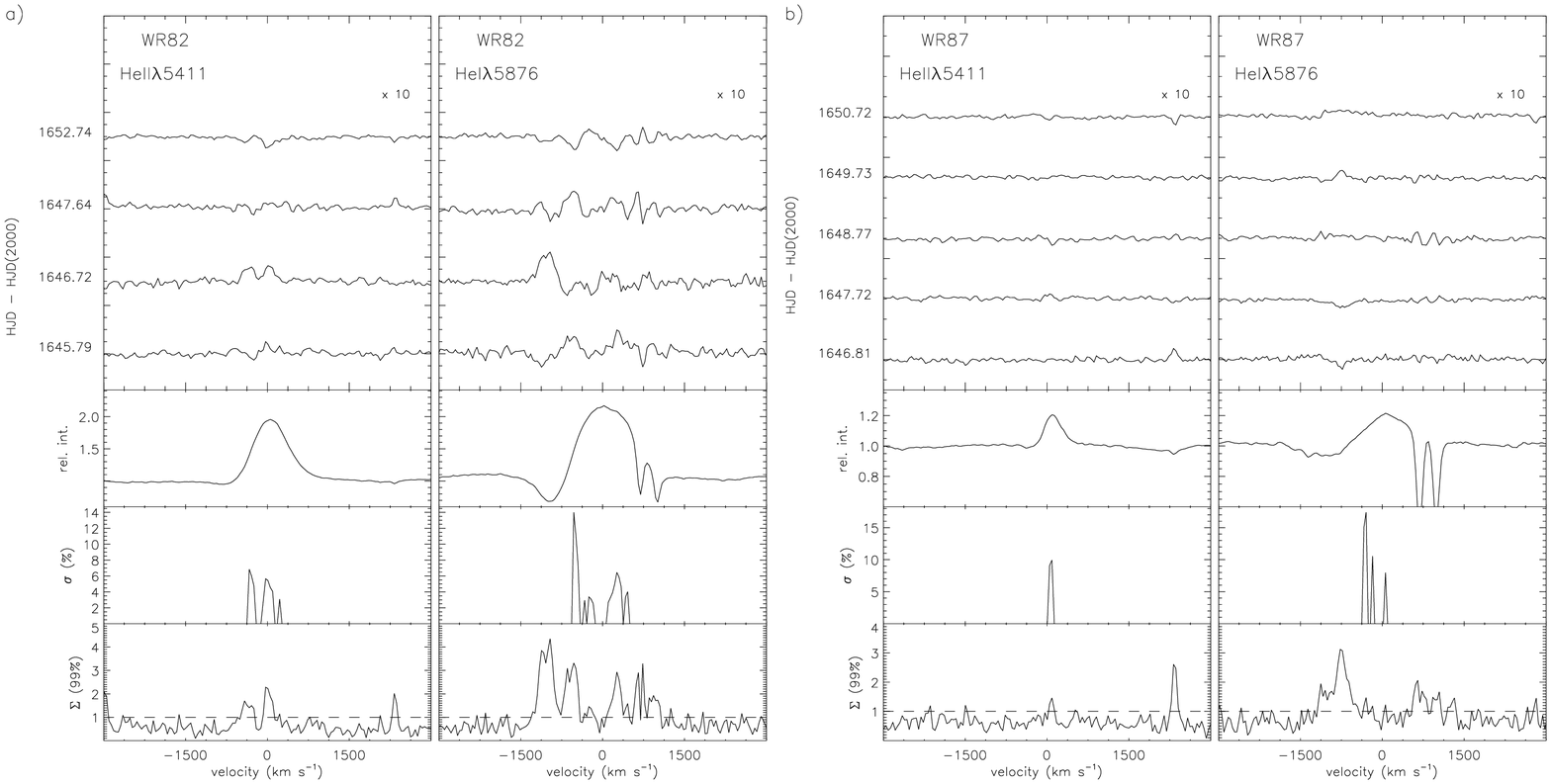}}
  \caption{Same as Figure~\ref{NordSud}(a) for a) WR\,82 (WN7(h)) and b) WR\,87 (WN7h+OB?).}
  \label{fig10S}
\end{figure}

\begin{figure}[tbp]
  \centerline{\plotone{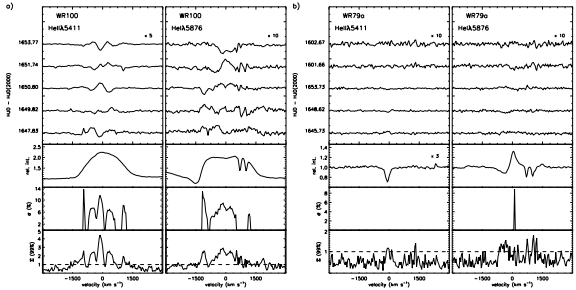}}
  \caption{Same as Figure~\ref{NordSud}(a) for a) WR\,100 (WN7) and b) WR\,79a (WN9ha).}
  \label{fig11S}
\end{figure}

\begin{figure}[tbp]
  \centerline{\plotone{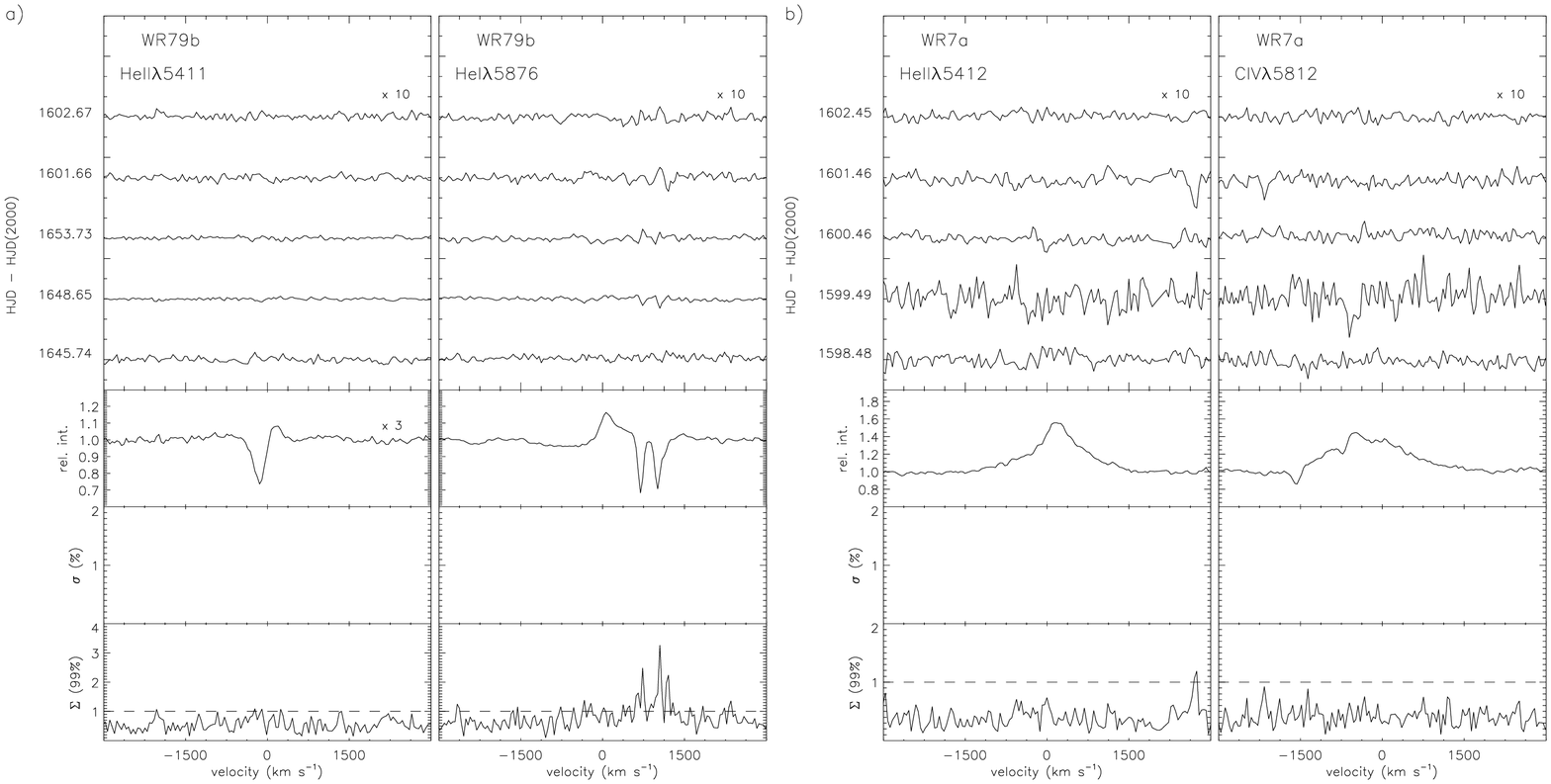}}
  \caption{a) Same as Figure~\ref{NordSud}(a) for WR\,79b (WN9ha). b) Same as Figure~\ref{NordSud}(a) for WR\,7a (WN4/WC), but for the He{\sc ii}$\lambda$5411 (left) and C{\sc iv}$\lambda\lambda$5802/12 (right) line-profiles.}
  \label{fig12S}
\end{figure}

\begin{figure}[tbp]
  \centerline{\plotone{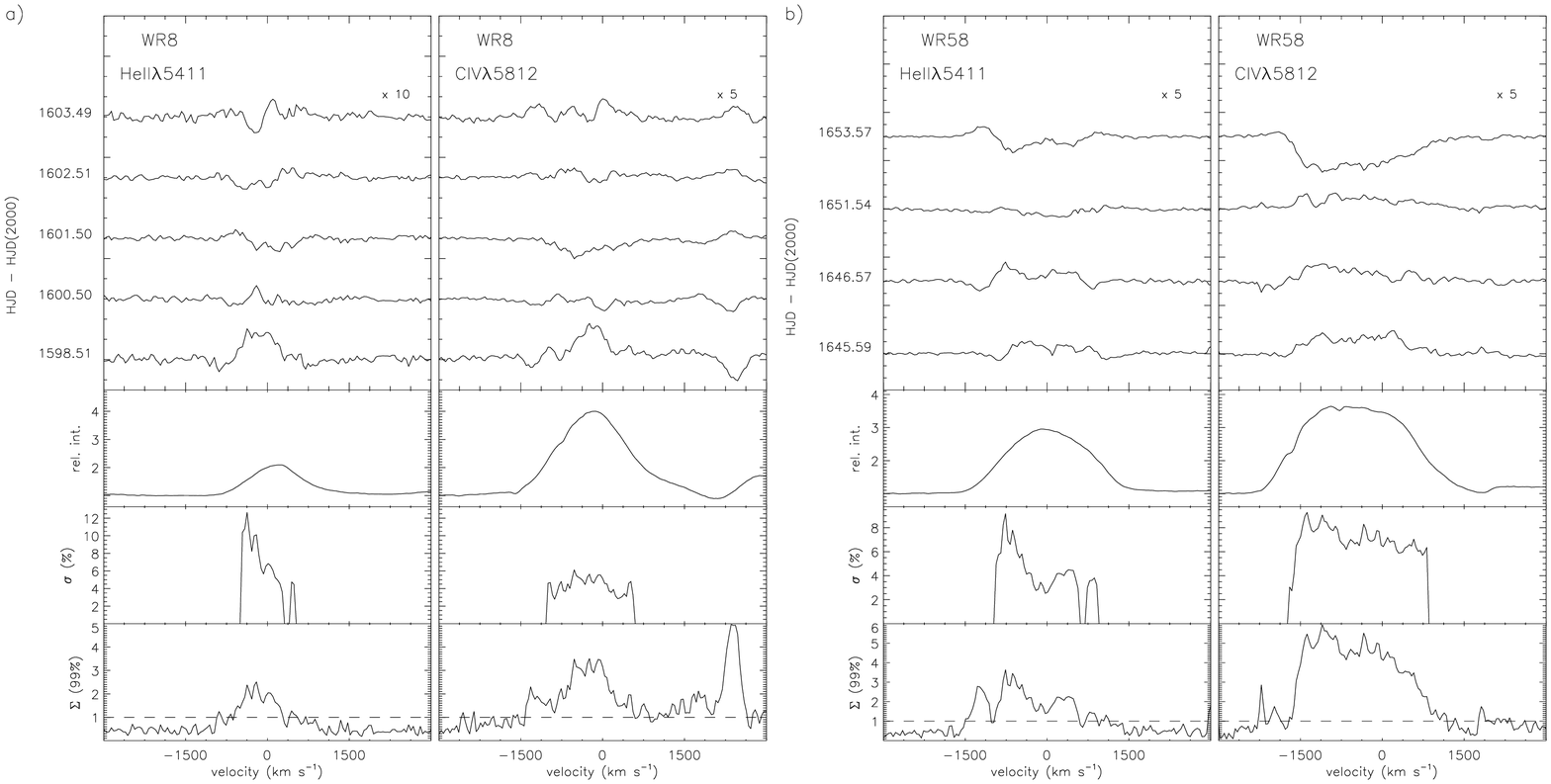}}
  \caption{Same as Figure~\ref{NordSud}(a) for a) WR\,8 (WN7/WCE+?) and b) WR\,58 (WN4/WCE), but for the He{\sc ii}$\lambda$5411 (left) and C{\sc iv}$\lambda\lambda$5802/12 (right) line-profiles.}
  \label{fig13S}
\end{figure}

\begin{figure}[tbp]
  \centerline{\plotone{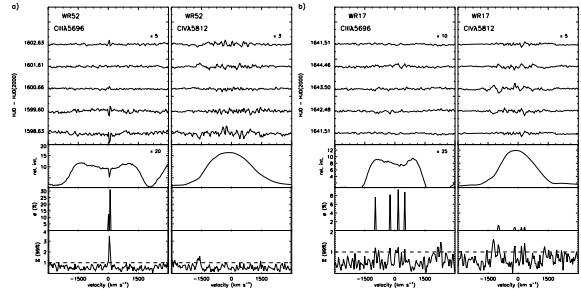}}
  \caption{Same as Figure~\ref{NordSud}(b) for a) WR\,52 (WC4) and b) WR\,17 (WC5).}
  \label{fig14S}
\end{figure}

\begin{figure}[tbp]
  \centerline{\plotone{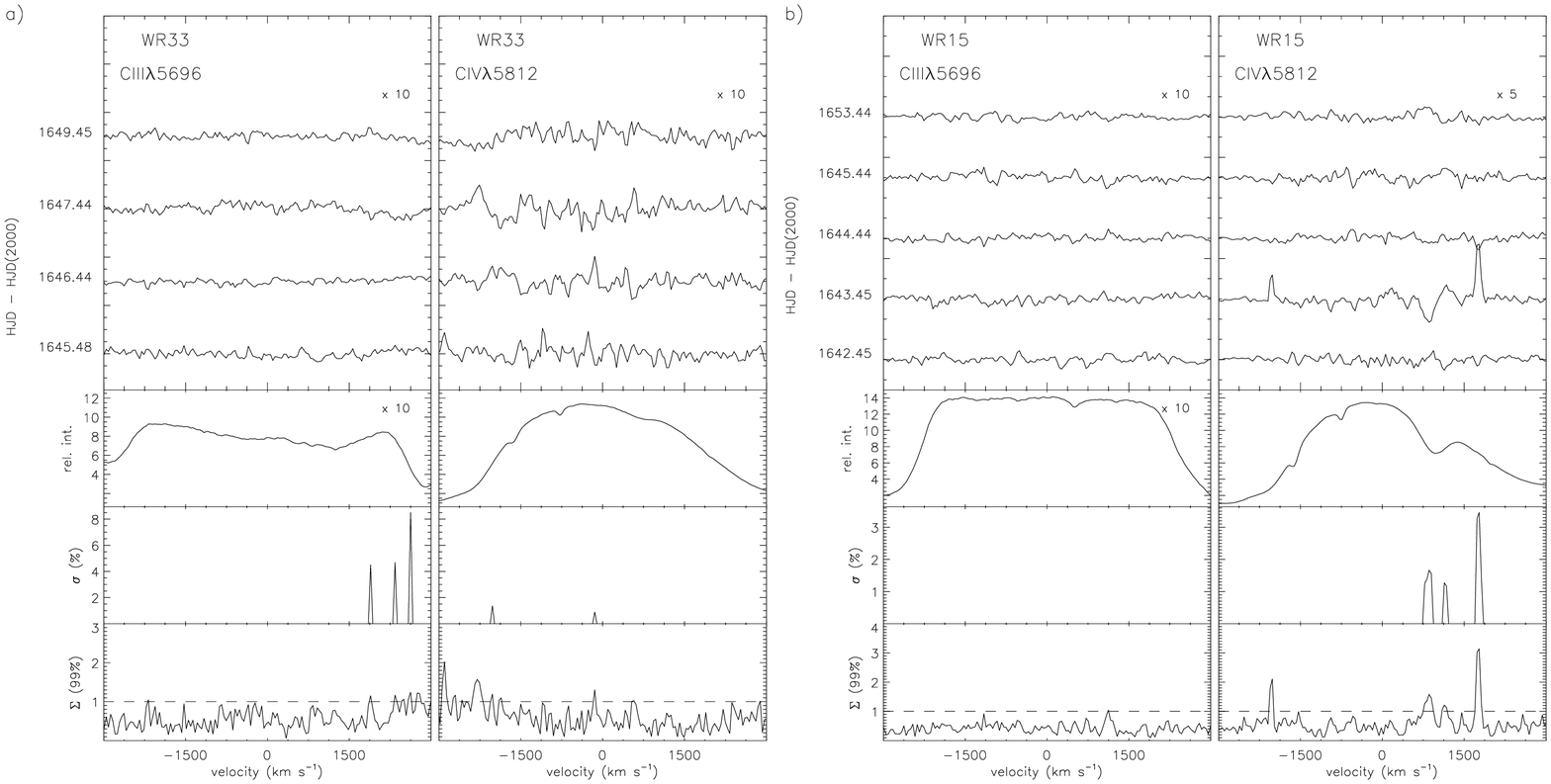}}
  \caption{Same as Figure~\ref{NordSud}(b) for a) WR\,33 (WC5) and b) WR\,15 (WC6).}
  \label{fig15S}
\end{figure}

\begin{figure}[tbp]
  \centerline{\plotone{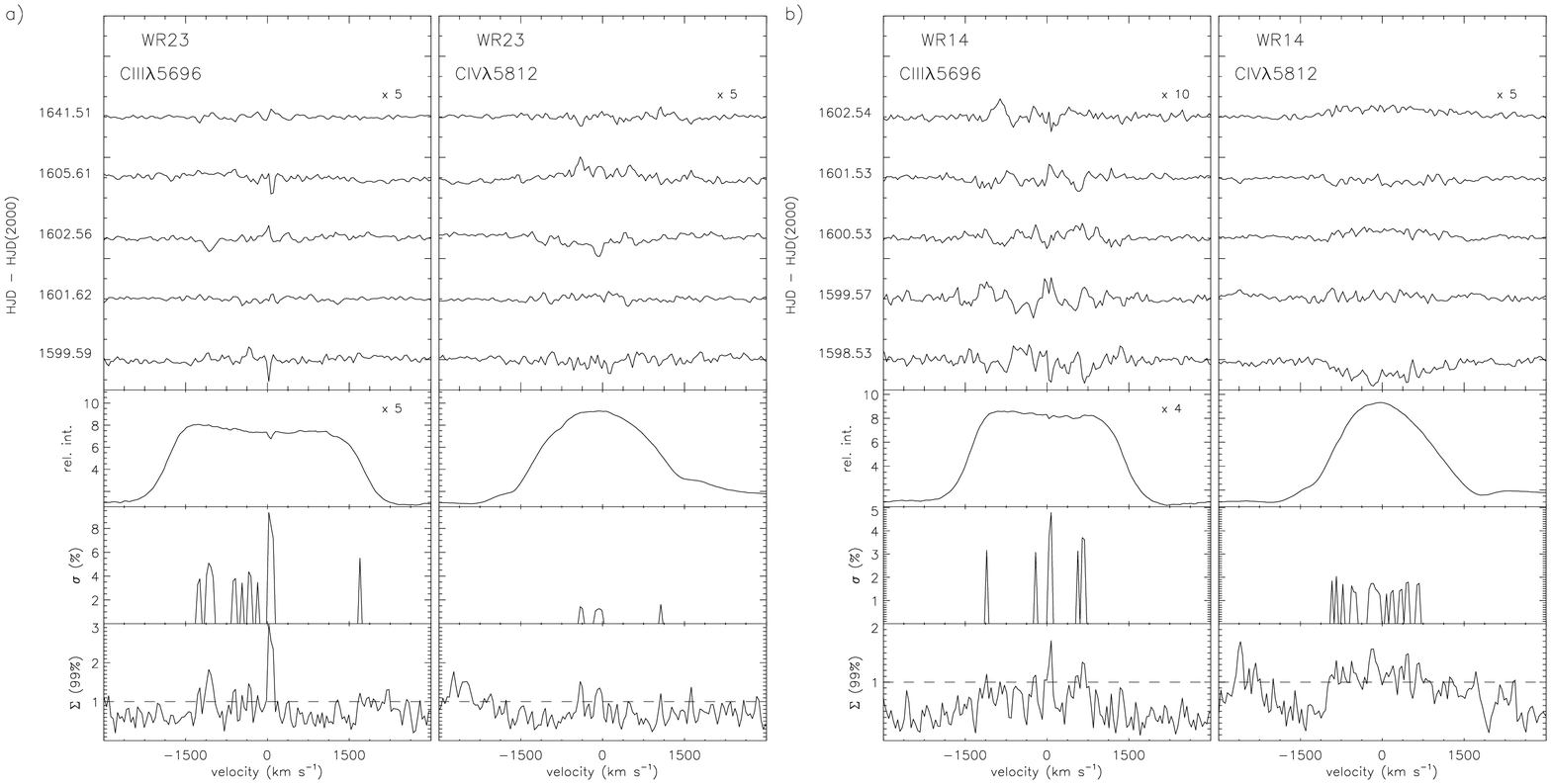}}
  \caption{Same as Figure~\ref{NordSud}(b) for a) WR\,23 (WC6) and b)  WR\,14 (WC7+?).}
  \label{fig16S}
\end{figure}

\begin{figure}[tbp]
  \centerline{\plotone{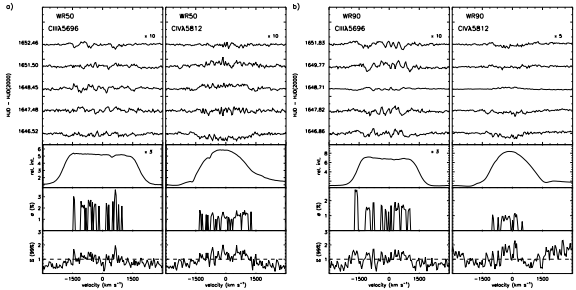}}
  \caption{Same as Figure~\ref{NordSud}(b) for a) WR\,50 (WC7+OB?) and b) WR\,90 (WC7).}
  \label{fig17S}
\end{figure}

\begin{figure}[tbp]
  \centerline{\plotone{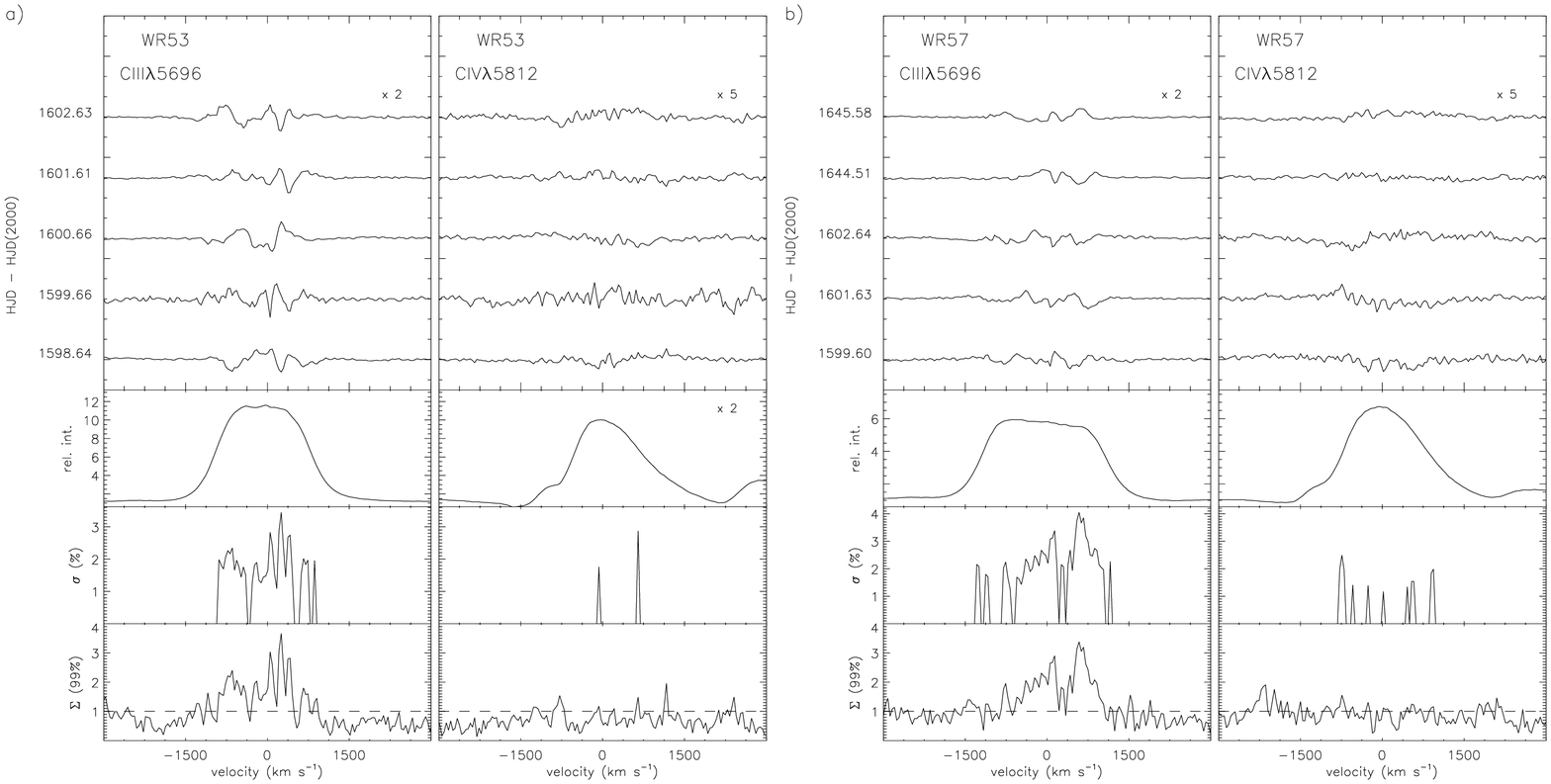}}
  \caption{Same as Figure~\ref{NordSud}(b) for a) WR\,53 (WC8d) and b) WR\,57 (WC8).}
  \label{fig18S}
\end{figure}

\begin{figure}[tbp]
  \centerline{\plotone{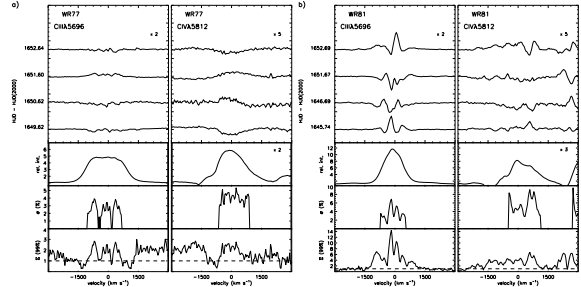}}
  \caption{Same as Figure~\ref{NordSud}(b) for a) WR\,77 (WC8+OB?) and b) WR\,81 (WC9).}
  \label{fig19S}
\end{figure}

\begin{figure}[tbp]
  \centerline{\plotone{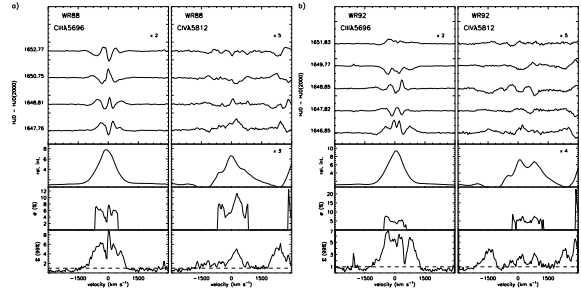}}
  \caption{Same as Figure~\ref{NordSud}(b) for a) WR\,88 (WC9) and b) WR\,92 (WC9).}
  \label{fig20S}
\end{figure}

\begin{figure}[tbp]
  \centerline{\plotone{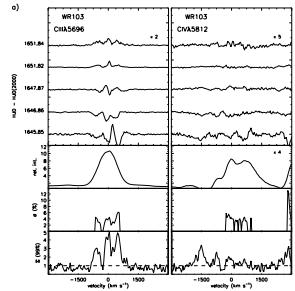}}
  \caption{Same as Figure~\ref{NordSud}(b) for a) WR\,103 (WC9d).}
  \label{fig21S}
\end{figure}

Figures~\ref{fig2S} to \ref{fig21S} present montages of residuals for all our targets. The graphs are organized by spectral type starting with WN stars from early to late types (Figures~\ref{fig2S} to \ref{fig12S}) followed by WN/WC stars (Figures~\ref{fig12S} and \ref{fig13S}) and finally WC stars from early to late types (Figures~\ref{fig14S} to \ref{fig21S}). In each of these figures, the $\sigma$-spectrum and the $\Sigma$ (99\%) spectrum are presented in the middle and the bottom panels respectively. In the following subsections, we will discuss in turn the variability for stars of different spectral types. Our conclusions concerning the spectroscopic variability level of each star is given in the last column of Table~\ref{Targets}

\begin{deluxetable}{ccccccccc}
\tabletypesize{\scriptsize}
\tablecolumns{9}
\tablewidth{0pc}
\tablecaption{Our Sample of Single WR Stars in the Southern Hemisphere}
\tablehead{
\colhead{Name} & \colhead{Spectral Type\tablenotemark{1}} & \colhead{RA (2000)} & \colhead{DEC (2000)} & \colhead{$v$} & \multicolumn{2}{c}{$v_{ \infty}$(km\,s$^{-1}$)}& \colhead{Variability}  }
\startdata
WR7a  & WN4/WC     &  07 20 22.38 &  -23 43 57.6 &  (12.5)& & & NV \\
WR8   & WN7/WCE+?  &  07 44 58.22 &  -31 54 29.6 &  10.48 & 1505\tablenotemark{2} & & LSV\\
WR10  & WN5ha      &  07 59 46.25 &  -28 44 03.1 &  11.08 & 1375\tablenotemark{2} & 1100\tablenotemark{3} & NV\\
WR14  & WC7+?      &  08 54 59.17 &  -47 35 32.7 &   9.42 & 2095\tablenotemark{2} & 2900\tablenotemark{4} & SSV\\
WR15  & WC6        &  09 13 11.77 &  -50 06 25.6 &  11.72 & 3600\tablenotemark{4} & & NV \\
WR17  & WC5        &  10 10 31.92 &  -60 38 42.4 &  11.03 & 2200\tablenotemark{4} & & SSV\\
WR18  & WN4        &  10 17 02.28 &  -57 54 46.9 &  11.11 & 1800\tablenotemark{3} & & NV \\
WR21a & WN6+O/a    &  10 25 56.49 &  -57 48 44.4 &  12.80 & & & NV \\
WR23  & WC6        &  10 41 38.33 &  -58 46 18.8 &   9.67 & 2240\tablenotemark{2} & 2900\tablenotemark{4} & SSV\\
WR24  & WN6ha      &  10 43 52.27 &  -60 07 04.0 &   6.49 & 1985\tablenotemark{2} & 2160\tablenotemark{3} & NV \\
WR28  & WN6(h)+OB? &  10 48 58.68 &  -59 03 37.5 &  12.98 & 1200\tablenotemark{3} & & SSV\\
WR33  & WC5        &  11 00 00.72 &  -57 48 59.5 &  12.35 & 4500\tablenotemark{4} & & NV \\
WR44  & WN4+OB?    &  11 16 57.86 &  -59 26 24.0 &  12.96 & 1400\tablenotemark{3} & & LSV\\
WR50  & WC7+OB     &  13 18 01.07 &  -62 26 04.5 &  12.49 & 3200\tablenotemark{4} & & SSV\\
WR52  & WC4        &  13 18 28.00 &  -58 08 13.6 &   9.86 & 2765\tablenotemark{2} & 3600\tablenotemark{4} & NV \\
WR53  & WC8d       &  13 30 53.26 &  -62 04 51.8 &  10.88 & 1700\tablenotemark{4} & & SSV\\
WR54  & WN5        &  13 32 43.79 &  -65 01 27.9 &  12.99 & 1500\tablenotemark{3} & & SSV\\
WR55  & WN7        &  13 33 30.13 &  -62 19 01.2 &  10.87 & 1200\tablenotemark{3} & & LSV\\
WR57  & WC8        &  13 43 16.37 &  -67 24 04.9 &  10.02 & 1750\tablenotemark{2} & 2100\tablenotemark{4} & SSV\\
WR58  & WN4/WCE    &  13 49 04.52 &  -65 41 56.0 &  13.05 & 1480\tablenotemark{2} & & LSV\\
WR61  & WN5        &  14 13 03.53 &  -65 26 52.7 &  12.41 & 1400\tablenotemark{3} & & LSV\\
WR63  & WN7+OB     &  14 50 58.31 &  -59 51 26.7 &  12.83 & 1700\tablenotemark{3} & & LSV\\
WR67  & WN6+OB?    &  15 15 32.61 &  -59 02 30.8 &  12.12 & 1500\tablenotemark{3} & & SSV\\
WR71  & WN6+OB?    &  16 03 49.35 &  -62 41 35.8 &  10.23 & 1200\tablenotemark{3} & & SSV\\
WR75  & WN6        &  16 24 26.23 &  -51 32 06.1 &  11.23 & 2300\tablenotemark{3} & & SSV\\ 
WR77  & WC8+OB     &  16 41 19.12 &  -48 01 59.6 &  13.00 & 2300\tablenotemark{4} & & SSV\\
WR78  & WN7h       &  16 52 19.25 &  -41 51 16.2 &   6.61 & 1335\tablenotemark{2} & 1385\tablenotemark{3} & SSV\\
WR79a & WN9ha      &  16 54 58.51 &  -41 09 03.1 &   5.29 & & & SSV\\
WR79b & WN9ha      &  16 55 06.45 &  -44 59 21.4 &   8.32 & & & NV \\
WR81  & WC9        &  17 02 40.39 &  -45 59 15.5 &  12.71 & 1300\tablenotemark{4} & & LSV\\
WR82  & WN7(h)     &  17 04 04.61 &  -45 12 15.0 &  12.41 & 1100\tablenotemark{3} & & SSV\\
WR83  & WN5        &  17 10 54.50 &  -46 36 18.0 &  12.79 & & & SSV\\
WR85  & WN6h+OB?   &  17 14 27.13 &  -39 45 47.0 &  10.60 & 1400\tablenotemark{3} & & SSV\\
WR87  & WN7h+OB    &  17 18 52.89 &  -38 50 04.5 &  12.59 & 1400\tablenotemark{3} & & SSV\\
WR88  & WC9        &  17 18 49.50 &  -33 57 39.8 &  13.25 & 1500\tablenotemark{4} & & LSV\\
WR90  & WC7        &  17 19 29.90 &  -45 38 23.8 &   7.45 & 2100\tablenotemark{2} & 1800\tablenotemark{4} & SSV\\
WR92  & WC9        &  17 25 23.15 &  -43 29 31.9 &  10.43 & 1130\tablenotemark{2} & 1300\tablenotemark{4} & LSV\\
WR100 & WN7        &  17 42 09.77 &  -32 33 24.7 &  13.44 & 1600\tablenotemark{3} & & LSV\\
WR103 & WC9d+?     &  18 01 43.14 &  -32 42 55.2 &   8.86 & 1060\tablenotemark{2} & 1300\tablenotemark{4} & SSV\\
\enddata
\tablenotetext{1}{Spectral types are from \citet{Huc01}}
\tablenotetext{2}{\citet{Prinja90}}
\tablenotetext{3}{\citet{Ham06}}
\tablenotetext{4}{\citet{Tor86}}
\label{Targets}
\end{deluxetable}

\clearpage

\subsection{WN stars}

A total of 5 WN stars are designated as NV. In the case of WR\,10, WR\,21a, WR\,24 and WR\,79b this is simply because the emission lines that are present in the selected wavelength range are too weak (or simply absent) and no variability can be detected with the data quality in hand. However, in the case of WR\,18 (and to a lesser extent WR\,10), in spite of the fact that the He{\sc ii}$\lambda$5411 line is quite strong, there is clearly no significant variability detected, excluding a few narrow peaks in the $\Sigma$-spectrum which barely exceed 1.

All other WN stars are variable at either a large or small relative level. Interestingly, most of these stars, if not all, show significant variability in the absorption component of the He{\sc i}$\lambda$5876 P~Cygni profile. Moreover, the deeper the absorption, the higher the level of the $\Sigma$-spectrum. This feature is in fact the only variable part of the spectrum for WR\,28 and WR\,79a, thereby allowing us to classify them as SSV and not NV. Since it is not possible to clearly separate the absorption and the emission components of P~Cygni profiles, and due to blending with the C{\sc iv}$\lambda\lambda$5802/12 line, it is not possible to determine the exact depth of the absorption component of the He{\sc i}$\lambda$5876 line, and no relevant value of $\sigma$ can be computed. Therefore $\sigma_j$=0 has been imposed at all wavelengths $j$ corresponding to an absorption in the mean profile. Note that WR\,18 and WR\,79b, both classified NV, do not show a clear P~Cygni absorption in the He{\sc i} line. This is probably due to strong blending with the C{\sc iv}$\lambda\lambda$5802/12 line in the case of WR\,18. The presence of stronger variability in the absorption components relative to the emission components of P~Cygni profiles has already been mentioned in Paper~I. We believe this is because the absorption part of the line profile  comes from a much smaller volume of the wind than the emission part and  therefore suffers less from cancelling effects from large-scale changes in  density or ionisation arising in different parts of the wind.

In addition to WR\,28 and WR\,79a, 9 other WN stars are classified as SSV. For WR\,54, WR\,67, WR\,71, WR\,75, WR\,82, WR\,83 and WR\,85, variability has been detected at a level of 3--5~\% of the line flux. The nature of the variability observed in these stars is similar to what was described in Paper~I in the case of SSV stars. These changes are associated with clumping in the WR wind, as observed for example by \citet{Mofa88} and \citet{Lepi96}. The nature of the changes is less clear for WR\,78 and WR\,87. Indeed, in those cases, the $\Sigma$(99\%)-spectrum shows only a few narrow peaks which are not very significant and the $\sigma$-spectrum has different amplitudes in the He{\sc i} and He{\sc ii} lines. This is due to the low intensity of the emission lines that do not exceed 1.3 times the continuum. In the case of such a low-intensity line, $(\bar{S}_j-1)$ takes values close to 0 and, when the variability has an amplitude as low as 2--3~\% of the line intensity, the noise level constitutes a non-negligible fraction of the $\sigma$-spectrum. Consequently, $\sigma$ reaches artificially high values. Note that for WR\,71 and WR\,75 the value of $\sigma$ reaches an amplitude of $\sim$5~\% of the line intensity which we adopted as the approximate limit between a SSV and a LSV classification. In this case, we decided to classify them SSV since the $\sigma$-spectrum is similar to what is observed for WR110, designated as SSV in Paper~I.

When the line intensity is sufficiently high, large-scale variability levels are observed in WR\,44, WR\,55, WR\,61, WR\,63 and WR\,100. In all cases, we see a highly significant level of variability that reaches $\sim$10~\% of the line intensity with a $\sigma$-spectrum that resembles the one presented in Paper~I for WR\,134, a well-known CIR-type variable \citep{Mor2}.

\subsection{WN/WC stars}

In our sample, only three stars have a hybrid WN/WC spectral type. WR\,7a, is classified as NV because no significant variability is seen in the $\Sigma$(99\%)-spectrum calculated for the moderately intense lines. WR\,8 and WR\,58 are classified LSV. Interestingly, the spectral lines of WR\,58 have a globally variable intensity as can be seen from the very broad features in the residuals. This is probably due to variability in the continuum which translates as a variable dilution level of the spectral lines, since we have normalized the continuum to one. Both WR\,8 and WR\,58 present a large-scale variability similar to what is expected for CIR-type variables.

\subsection{WC stars}

For WR\,15, WR\,33 and WR\,52, no significant variability is observed, even in very strong lines, and therefore, we classify them as NV. As for WR\,14, WR\,17, WR\,23, WR\,50, WR\,53, WR\,57, WR\,77, WR\,90 and WR\,103, the variability detected reaches a level of $\sigma$=1--4~\% of the line intensity and therefore we classify them as SSV. Note that the high value of $\sigma$ in the C{\sc iii} line of WR\,17 is disregarded in view of its low intensity (see above). Also, without the normalization of the line intensity described above in section~\ref{VarSer}, WR\,77 would show a much higher level of variability due to what seems to be real changes in the continuum. This is not the type of structure we are looking for and we prefer to keep it has SSV, but it would still be an interesting case to investigate. 

The only stars classified as LSVs are three WC9s out of four in our sample. Indeed, they reach a level of variability of more than 5~\% of the line intensity. However, as described in Paper~I, WC9 stars are possible dust producers although in this case they have not yet been identified as such. We will come back to this point in the next section.

\section{Discussion and Conclusions}\label{Disc}

This study completes our systematic search for Corotating Interaction Regions candidates among the brightest, presumably single, Galatic WR stars.  In this section, we present a global summary of the results for the sample presented here, combined with those obtained for our northern sample and presented in Paper~I.

\begin{figure}[tbp]
 \centerline{\plotone{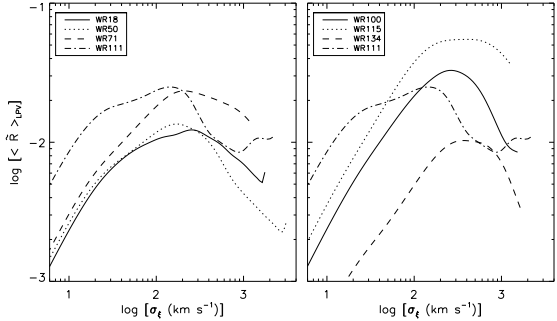}}
  \caption{Mean wavelet power spectrum $\langle\tilde{R}(\sigma_\xi)\rangle_{\rm LPV}$  for the spectral time series of seven stars in our sample. In the left panel are shown our results for WR\,18 (NV), WR\,50, WR\,71 and WR\,111 (SSV), which show a maximum at $\bar{\sigma}_\xi\sim100\,km\,s^{^-1}$. In the right panel are presented the stars WR\,100, WR\,115 and WR\,134 (LSV), which show a maximum at $\bar{\sigma}_\xi\sim300-500\,km\,s^{^-1}$. $\langle\tilde{R}(\sigma_\xi)\rangle_{\rm LPV}$ for WR\,111 was repeated in the right panel for better comparison.}
  \label{fig22S}
\end{figure}

\begin{figure}[tbp]
 \centerline{\plotone{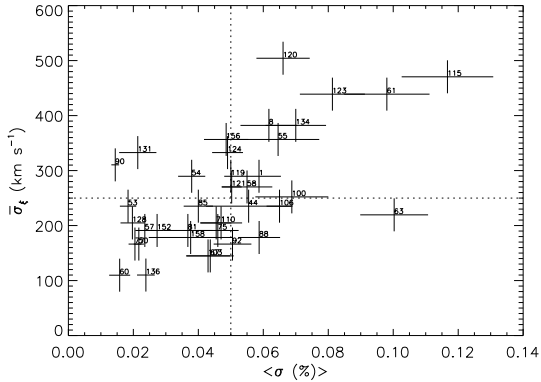}}
  \caption{Maximum of the mean wavelet power spectrum, $\bar{\sigma}_\xi$, of the stars of our data set as a function of $\langle\sigma(\%)\rangle$, i.e. the mean value of their $\sigma$-spectrum. Stars for which no $\sigma$-spectrum could be computed have been removed. For each star, marked in the plot by its WR number, a 1$\sigma$ error bar is plotted in both axis. The vertical dotted line marks the limit of $\sigma > 5\%$ adopted in this study, and the vertical one marks $\bar{\sigma}_\xi > 250\,km\,s^{-1}$. All LSV stars are in the upper-right quadrant, except for WR\,63 (see text).}
  \label{fig23S}
\end{figure}

Following the method we have used in Paper~I, we based our distinction between the different types of variability solely on the level of variability. However, we could also use the velocity dispersion of the structures in the residuals. Indeed, \citet{Le99} define $\langle\tilde{R}(\sigma_\xi)\rangle_{\rm LPV}$, i.e. the mean wavelet power spectrum of the residuals averaged over the spectral domain $\xi$, to determine the level of variability. The maximum of this power spectrum, $\bar{\sigma}_\xi$, gives the mean line-of-sight velocity dispersion of line profile variability subpeaks. We calculated $\langle\tilde{R}(\sigma_\xi)\rangle_{\rm LPV}$ following the formalism of \citet{Le99} for our complete sample of stars. In Figure~\ref{fig22S}, we present our results for WR\,18 (NV), WR\,50, WR\,71 and WR\,111 (SSV) and WR\,100, WR\,115 and WR\,134 (LSV). The profiles for WR\,111 and WR\,134 are very similar to those in Figure~4 of \citet{Le99} for the same stars, but with a spectral resolution more than ten times lower and a signal-to-noise of 100 instead of 200, we do not resolve as well structures with velocity dispersion lower than $40\,km\,s^{-1}$. Moreover, the noise pattern can sometimes mimic true signal at a few hundreds $km\,s^{-1}$, and 5 spectra is not enough to average this artifact out from the mean wavelet power spectrum. To minimize these effects, we filter out any structures smaller than $10\,km\,s^{-1}$ and bigger than the terminal velocity $v_\infty$ from the residuals. However, in the cases of low or very low level of variability, the mean wavelet power spectrum reaches the maximum at very large scales, which is meaningless when no significant variability is detected (e.g. WR\,18 for which we get $\bar{\sigma}_\xi\sim300\,km\,s^{^-1}$). From Figure~\ref{fig22S}, we get  $\bar{\sigma}_\xi\sim 100\,km\,s^{-1}$ for the SSV cases and $\bar{\sigma}_\xi = 300-500\,km\,s^{-1}$ for the LSV cases. To verify if the distinction between the different types of variability is equivalent, regardless of the method used, we plot in Figure~\ref{fig23S} the $\bar{\sigma}_\xi$-value of the stars in our sample as a function of $\langle\sigma(\%)\rangle$, i.e. the mean value of their $\sigma$-spectrum. Dotted lines mark the limit of $\sigma > 5\%$ adopted in this study, as well as $\bar{\sigma}_\xi > 250\,km\,s^{-1}$, which seems to separate SSV to LSV cases in both this study and \citet{Le99}. Obviously, all the stars for which no good values of $\sigma(\%)$ could be computed due to a too low level of variability are excluded from the plot. It is clear from this figure that stars with a high mean line-of-sight velocity dispersion also have a high level of variability and that LSV are mostly located in the upper right quadrant of the plot. Some cases seem to disagree with the general trend. Indeed, WR\,90, WR\,131 and WR\,54 show variations with a level of variability significantly lower than 5\%, but with a mean velocity dispersion higher than $250\,km\,s^{-1}$. But after inspection, we find that this is caused by the small number of spectra, since the stochastic appearance of clumps on the spectral lines combined with the 
noise pattern giveS the impression of shallow large-scale bumps on the line profile and these are clearly different from the structures we are looking for. As for WR\,63, high $\sigma$-values are obtained, even if the mean velocity dispersion of the variabilities is low. This is either due to the low intensity of the observed spectral line combine with a fairly high variability level (but still lower than 5\%), which artificially boosts the $\sigma$-value, or to the presence of real structures in the wind that actually have a small mean velocity dispersion value. Because we cannot distinguish the two possibilities, we still keep WR\,63 in our list of LSVs, but we must keep in mind that this may not be the best CIR-type variable candidate. Finally, we conclude that the determination of the type of variability is still more robust and easier to obtain when based on the level of variability, since this method is less dependent on the signal-to-noise ratio or on the spectral resolution than the method using the mean velocity dispersion of the variabilities.

\begin{figure}[tbp]
  \centerline{\plotone{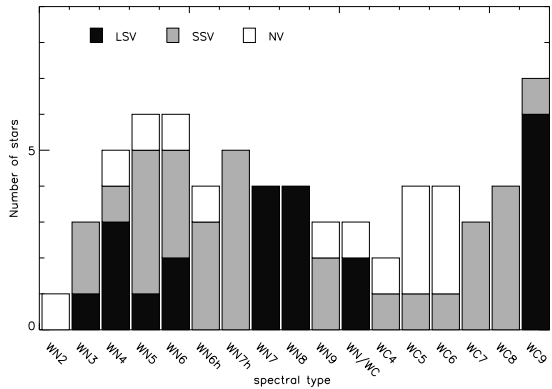}}
  \caption{Histogram of the spectral type of the 68 studied
stars. This includes the 25 stars discussed in Paper~I, the 39 stars
presented in this paper and 4 stars for which spectra has already been
published in the literature, i.e.  WR\,6 \citep{Mor1}, WR\,40
\citep{Ant}, WR\,46 \citep{Veen} and WR\,138 \citep{Lepi96}. Marked in
black are the stars identified as LSV, in grey the stars we classified
as SSV, and the NV are in white.}
  \label{fig24S}
\end{figure}

A compilation of the variability diagnostics as a function of the spectral type is presented in Figure~\ref{fig24S}. We add to this sample data from the literature for the stars WR\,6 \citep{Mor1}, WR\,40 \citep{Ant}, WR\,46 \citep{Veen} and WR\,138 \citep{Lepi96}. 

Among WNE stars (including WN6 stars without hydrogen), 4 are classified NV, 10 SSV and 8 LSV. Among WNL stars (including WN6 stars with hydrogen), 2 are NV, 10 SSV and 8 LSV. Interestingly, all the WN7 stars without hydrogen present in our sample are, with the WN8 stars, the only WNL stars classified as LSV. Due to the proximity of the two spectral types, it remains to be seen if the variability observed in WN7 stars is similar to the variability observed in WN8 stars and possibly related to pulsations. Only a better monitoring of these stars with a more intensive temporal sampling than that obtained for this study can provide more clues towards answering this question.

Concerning spectral variability, we find no clear difference in the fraction of stars we find in the different categories among WNE and WNL stars. Indeed, we find 18\% 45\% and 36\% of NV, SSV and LSV respectively amoung WNE stars compared to 10\%, 50\% and 40\% of NV, SSV, LSV respectively amoung WNL stars. For stars showing small-scale variability, the numbers above may be misleading. Indeed, small-scale variations could be easily masked by large-scale changes. Therefore, it is impossible from the present study to provide an accurate fraction of stars showing this type of variability and the numbers above represent, in that case, only lower limits. For stars presenting large-scale variations, the difference between WNE and WNL stars is not in the fraction of stars we classified in the different categories, but rather in their distribution among the subtypes. Indeed, while among WNE stars almost all spectral subtypes show some stars with large-scale variability, within WNL stars there is a striking dichotomy: None of the 9 WN6h, WN7h stars show large-scale variability while all 8 WN7 and WN8 stars do.

Among WC stars, only those of spectral type WC9 show LSV. The only WC9 star which is not a LSV is WR\,103, which we classified as SSV and for which the $\sigma$-spectrum barely reaches the 5~\% level, i.e. just at the limit to be classified as LSV according to the criteria we adopted in Paper~I. It has been proposed in Paper~I that the high level of variability of WC9 stars could potentially be related to excess emission from a yet unidentified shock cone formed following the collision of two stellar winds in a binary system. Note that in principle, the stars in our sample are {\it not known} to be binaries and this remains to be checked via a more intensive spectroscopic monitoring. Interestingly, all the WC9 stars in our sample show a high level of variability, even if they are not known to be dust producers, such as WR\,81, WR\,88 and WR\,92. Moreover, the only dust-producer of WC8d subtype in our sample, WR\,53, is classified as SSV, since its $\sigma$-spectrum reaches only a level of 2 and 3~\%. Hence, we conclude that the large-scale variability is a common feature of all WC9 stars, but not necessarily of all dust-producers. What the origin of this variability is, either CIR-type or excess emission from a wind-wind collision region, remains to be identified. This can only be done through a more intensive spectroscopic observing campaign; only then can the kinematical signature of binary-related or CIR-type changes be recognized. All other WC stars are NV (7 WCE) or SSV (3 WCE and 7 WCL). None of the WCE stars in our sample shows large-scale spectral variability. Therefore, the determination of rotation periods from CIRs seems to be impossible for this type of star. The reason for this remains to be determined. However, this is unfortunate, since WCE stars are thought to be possible progenitors of the long-soft Gamma-Ray Bursts \citep{Woo93,Mac99,Mac01}. Even if CIRs were present in all WR winds, it is possible that they are not visible in all emission lines. For example, they may remain confined to regions close to the surface of the star in which case they may not be present in line-formation regions very far from the base of the wind, which is the case for the flat-topped C{\sc iii}$\lambda$5696 line. It is clear  from the different line shapes of C{\sc iii}$\lambda$5696 and C{\sc iv}$\lambda\lambda$5802/5812 that the former is formed further out in the wind than the latter. However, it may still be too far fro the CIRs to be deteted in this line.

Finally, two WN/WC stars, WR\,8 (WN7/WCE) and WR\,58 (WN4/WCE), are identified as LSV. In addition to its line-profile variability, WR\,58 shows a variable dilution in the spectral lines most likely due to variability in the continuum. More observations are needed to determine if these variations are caused by the orbit of a companion star, but we prefer to keep WR\,58 in our list of CIR-type variable candidates for now. The only other WN/WC star in our sample, WR7a (WN4/WC), is classified NV.

We wish to emphasize once more that stars that we have classified as NV are not necessarily constant; it is simply that no variability has been detected in spectra with a given noise level. A more stringent limit for the absence of variability can be established for some stars. For example, in the case of WR\,2, we have obtained 5 spectra with a typical SNR~=~120 from which we deduce from the $\Sigma$-spectrum that the variability level is lower or equal than $\sim$~0.8~\% of the He{\sc ii}$\lambda$4686 line intensity. Moreover, we cannot detect changes in WR\,2 if it has a timescale a lot shorter than the exposure time ($\sim$~30~mins) or greater than the time spanned by our time-series (96~days). This establishes a rough limit of detectability of variability in the case of WR\,2. Of course, due to our poor sampling, it is clearly not enough to cover all the possible periods/timescales in the time range span and all the 5 spectra could have been taken in a ``quiet'' phase of the star. In Table~\ref{NV}, we present upper limits for the spectroscopic variability for all NV stars in our sample, except for WR\,21a which does not show emission lines at all in the chosen spectral range. We list the star's name, the emission line chosen to carry out the analysis, the maximum relative amplitude of the variability compared to the line flux, and the total exposure time range spanned by the observations in each case. The line chosen here is the strongest, non-blended line available in the spectral range.

\begin{table}[ht]
 \centering
  \caption{Upper limits for the level of variability in NV stars}
  \begin{tabular}{@{}ccccc@{}}
  \hline
   WR  &      spectral line      & max. rel. &  exp. time  &  time range \\
         &                         &    amplitude     &   (min)  &     (days)    \\
   \hline
   2   & He{\sc ii}$\lambda$4686 &     0.8~\%       &  30  &  96  \\
   4   & C{\sc iv}$\lambda$5016  &     1.2~\%       &  30  &  120 \\
   5   & C{\sc iv}$\lambda$5016  &     1.2~\%       &  30  &  92  \\
   10  & He{\sc ii}$\lambda$5412 &     1.5~\%       &  30  &  5   \\
   15  & C{\sc iii}$\lambda$5696 &     1.2~\%       &  30  &  9   \\
   18  & He{\sc ii}$\lambda$5412 &     1.2~\%       &  30  &  40  \\
   33  & C{\sc iii}$\lambda$5696 &     1.3~\%       &  30  &  4   \\
   52  & C{\sc iii}$\lambda$5696 &     1.5~\%       &  30  &  5   \\
   111 & C{\sc iii}$\lambda$5696 &     1.1~\%       &  10  &  30  \\
   154 & C{\sc iv}$\lambda$5016  &     1.2~\%       &  30  &  483 \\
   \hline
  \end{tabular}
  \label{NV}
\end{table}

From all LSV stars identified, we have established a list of promising new CIR-type variable candidates. Of course, from this list, we removed all the WN8 and WC9 stars, since it is possible that they are highly variable for other reasons than the presence of CIRs in their wind. Therefore, the 10 remaining candidates are WR\,1 (WN4), WR\,8 (WN7/WCE), WR\,44 (WN4), WR\,55 (WN7), WR\,58 (WN4/WCE), WR\,61 (WN5), WR\,63 (WN7), WR\,100 (WN7), WR\,115 (WN6) and WR\,120 (WN7). Including the two already known cases, WR\,6 (WN4) and WR\,134 (WN6), we get 12 CIR-type variable stars out of a total of 68 sampled WR stars (i.e. $\sim$ 18\%). If we exclude the 10 WN8 and WC9 stars from the total number of stars in the sample, this fraction of CIR-type variables becomes 12/58~$\sim$~21\%, a fraction which is equivalent to the fraction already determined in Paper~I, but now based on a more complete sample. Note that the WN4h star WR\,46 also shows clear, rapid, large-scale variability in photometry and spectroscopy. X-rays observations do not contradict the possibility of the presence of CIRs \citep{Gos11}, however intense monitoring campaigns have led \citet{Veen}, \citet{Oli04} and \citet{Hen10} to conclude that the origin of the changes is most likely non-radial pulsations. Indeed, for that star, the period determined in the various datasets, although always of the same order of magnitude ($\sim$8~hours), is not always strickly the same. This is highly incompatible with a rotation origin of the variability.

Using the stellar and atmospheric parameters of our 10 new CIR-type variable candidates determined by \citet{Ham06} from a quantitative spectral analysis using their Potsdam WR blankeded atmosphere model, we attempted to determine if these stars present any particular physical characteristic that could reveal information on the nature or origin of this type of structure in the wind. To do so, we have compared the distribution histograms of our 10 candidates to that of the entire \citet{Ham06} sample for the following parameters~: mass-loss rate, wind terminal velocity, luminosity, radius and temperature. We performed a cross-correlation between the two distributions to check for differences and nothing striking was found, although our results are based on only 10 candidates. Ideally, this should be verified once a larger CIR candidates sample has been identified.

The next step is to determine whether our candidates really show CIRs. To do this, we have initiated more intensive monitoring of our targets in photometry and in spectroscopy on different timescales ranging from 10~mins to several days over a long period of time in order to search for periods in the large-scale spectral variations. To date, we have obtained several nights on different telescopes to monitor WR\,1 \citep{Che10}, WR\,115 and WR\,120. Also, we must keep in mind that, even if we classify them as SSV, WR\,71 (WN6), WR\,75 (WN6) and WR\,110 (WN5-6) also show interesting $\sigma$-spectra, and they have to be considered as second priority targets. Finally, we believe it would be extremely useful to extend the sample of observed stars to fainter targets in order to identify more CIR candidates and thus generate a more statistically viable sample.

\acknowledgments
ANC gratefully acknowledges support from the Chilean {\sl Centro de Astrof\'\i sica} FONDAP No. 15010003 and the Chilean Centro de Excelencia en Astrof\'\i sica y Tecnolog\'\i as Afines (CATA). We also wish to thank the National Sciences and Engineering Research Council (NSERC) of Canada for financial support.

{\it Facilities:} \facility{CTIO}.

\clearpage

\end{document}